\documentclass[12pt]{article}
\usepackage{latexsym,amsmath,amssymb,amsbsy,graphicx}
\usepackage[utf8]{inputenc}
\usepackage[T2A]{fontenc}
\usepackage[space]{cite} % If exist.

\makeatletter\def\@biblabel#1{\hfill#1.}\makeatother
\allowdisplaybreaks
\begin {document}

\noindent\begin{minipage}{\textwidth}
\begin{center}

{\Large{New infrared camera of the Caucasian Mountain Observatory of the SAI MSU: design, main parameters, and first light}}\\[9pt]

{\large Zheltoukhov S.\,G.$^{1,2}$, Tatarnikov A.\,M.$^{1,2a}$, Belyakova A.\,A.$^2$, Koksharova E.\,A.$^2$}\\[6pt]

%\parbox{.96\textwidth}{\centering\small\it
\textit {$^1$ Faculty of Physics, M.V.Lomonosov Moscow State University, Moscow 119991, Russia.}\\
\textit {$^2$Sternberg Astronomical Institute, M.V.Lomonosov Moscow State University, Moscow 119191, Russia.}\\
\textit {E-mail: $^a$andrew@sai.msu.ru} \\ [1cc] 

%\parbox{.96\textwidth}{\centering\small Received 27.11.2023; Accepted 30.11.2023} 
\end{center}

{\parindent5mm This paper presents a prototype of an infrared photometer, created at SAI of MSU based on the commercial infrared module Gavin-615A. The operating spectral range of the photometer is  3-5~$\mu$m. Investigations of the photometer’s detector have shown that its parameters coincide with those stated by the manufacturer. The nonlinearity of the detector does not exceed $\sim5\%$ across the entire signal range, and coefficients for correction functions were determined. Additionally, we determined the readout noise $RN=1200\pm210$~e$^-$, the conversion coefficient GAIN~$=520\pm9$~e$^{-}/$ADU, the signal magnitude of the bias frame BIAS~$= 960.5 \pm 2.2$~ADU, and the dark current $\approx(9.3\pm1.1)\cdot10^6$~e$^-$/s, which is the sum of the detector’s dark current and the radiation from the entrance window of the detector module. The value of dark current was measured at a window temperature of  $6^\circ$C. Observations commenced with the photometer at the 2.5-m telescope of the Caucasian Mountain Observatory of MSU, with the first results presented in this paper. The unvignetted field of view was $30''$. In the $M$ band under good atmospheric conditions, an image quality close to the diffraction
limit was achieved. Images of a star with brightness $L=7.96$ and $M=6.78$ were obtained over the 20-second exposure time and a SNR$\sim10$ ratio. t is shown that at high image quality with SNR=3
and exposure of 20 seconds it is possible to observe stars up to $L\sim$9$^m$ and $M\sim$8$^m$. The main module of the photometer was also used in measurements of the sky background brightness.

\vspace{2pt}\par}


\vspace{1pt}\par

DOI: 10.55959/MSU0579-9392.79.2410801

\vspace{1pt}\par
\end{minipage}

% ======================================

\section*{Introduction}
\mbox{}\vspace{-\baselineskip}

The infrared range occupies almost 10 octaves on the electromagnetic waves scale, while the visible range, in which most astronomical observations are made~---just one octave. All types of astronomical objects~--- from exoplanets and active galactic nuclei to solar system bodies and artificial satellites can be observed in the infrared range. At the same time, observations in the IR range have a number of advantages: the maximum emission of cold objects (stars of late spectral types, dust envelopes, interstellar gas-dust clouds, etc.) falls in the IR region of the spectrum, low interstellar absorption allows us to observe objects hidden behind dense dust clouds, lower sensitivity to atmospheric turbulence makes it easier to obtain high angular resolution, and the sharp decrease in light scat
tering in the Earth’s atmosphere with increasing wave length allows for daytime observations in the mid-IR range.

The possibility of making observations of faint objects in the IR range depends largely on the location of the observatory and its astroclimate. Regular observations in this region of the spectrum are carried out in Russia in two observatories~--- the Crimean Astronomical Station of the SAI MSU with a single-element photometer based on an InSb photodiode (\cite{Shenavrin2011}) and the Caucasus Mountain Observatory of the SAI MSU (CMO, \cite{Shatsky2020}) with the ASTRONIRCAM spectrograph camera (\cite{Nadjip2017}) based on the Hawaii-2RG matrix detector. The astroclimate of CMO is well studied. This observatory is well suited for IR observations due to its high altitude (2110~m), low water vapor amount in the atmosphere (median PWV less than 8 mm \cite{Kornilov2016a}), and high image quality (\cite{Kornilov2014}). According to \cite{Tatarnikov_fon2023}, 1 square arcsec of the night sky has an average brightness of $J=15.5^m$, $H=13.7^m$, $K=13.1^m$, which corresponds to the best observatories of the world located at similar altitudes above sea level.

In recent years, significant progress has been made in the technology of IR detectors~--- the size of the sensitive region is increasing, quantum efficiency is close to unity, and the cost of some detector is decreasing. Estimates made in \cite{SZh2022} show that when using a ground-based telescope not optimized for IR observations, modern commercial matrix detectors can be used at wavelengths up to 5~µm~--- noise limitations in observations of faint objects will be determined mainly by background radiation, not by the noise of the detector itself.

At the end of 2022, MSU acquired a commercial IR matrix detector designed for thermal imaging cameras with an operating wavelength range of $3 - 5$~µm, on the basis of which we built a prototype of an astronomical IR camera, which was given the abbreviation LMP~--- L and M bands Photometer by the CMO and installed in 2023 on the 2.5-m telescope of this observatory. In this paper we present a characterization of the LMP camera as an instrument for astronomical observations in the $3-5$~µm wavelength region and the results of the first observations with it.

\section{Instrument design}
\mbox{}\vspace{-\baselineskip}

At present, the LMP photometer is assembled according to the scheme proposed earlier in our paper \cite{SZh2022}. It is not the most optimal scheme from the point of view of the instrumental background value, but it allows us to build a working instrument at minimal cost. In the future, we plan to add a focus reducer and move all the optical elements of the camera into a separate cooled module.

The photometer is based on a Gavin-615A light-sensitive module with a 640x512 HgCdTe matrix as an image detector cooled to $\approx82$~K by a Stirling machine. From the basic model of the module, the manufacturer, at our request, removed the internal bandpass filter. This made it possible to extend the spectral range of the camera sensitivity and to use external light filters realizing standard photometric bands coinciding with the windows of transparency of the Earth's atmosphere.

\begin{figure}[h]
\center{
\includegraphics[width=0.49\linewidth]{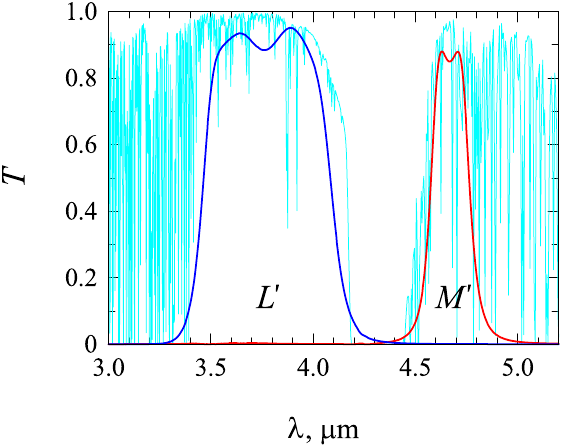}}
\caption{Transmission curves of  $L^\prime$  and  $M^\prime$ filters and the Earth’s atmosphere for an altitude of 2 km above sea level and a PWV=5 mm}
\label{fig:lm_atm}
\end{figure}

Initially (until November 2023), $L$ $(\lambda_{c}=3.7$~$\mu$m, $\Delta \lambda_{0.5}=0.49$~$\mu$m) and $M$ $(\lambda_{c}=4.8$~$\mu$m, $\Delta \lambda_{0. 5}=0.54$~$\mu$m) filters similar to those used in the single-channel infrared photometer on the 1.25-m ZTE telescope of the Crimean Astronomical Station \cite{Shenavrin2011}.  The expansion of the sensitivity range also allowed us to install the $K$ filter with a central wavelength of 2.2~$\mu$m, but the quantum efficiency of the camera at this wavelength is several times lower than in the main range, and only the brightest objects can be observed with it. As suggested in \cite{SZh2022}, mirrored surfaces are installed around the light filters to reduce the instrumental background from the filters. All the light filters, as well as the open aperture and calibration plug, are mounted in a wheel driven by a stepper motor. A spring-loaded stop mechanism (detend mechanism) is used to lock the wheel in the operating position. The filter change time is approximately 5~s.

In November 2023, we replaced the $L$ and $M$ filters with new ones manufactured by the Russian company ALKOR Technologies LLC\footnote{https://alkor.net/}, realizing the photometric bands $L^\prime$ and $M^\prime$ of the MKO-NIR (\cite{mko2}) system. Their response curves are shown in Fig.~\ref{fig:lm_atm}. They are chosen to fit optimally into the corresponding transparency windows of the Earth's atmosphere. The average transmission of the filters calculated by the formula $T_{avr}=\int T(\lambda)d\lambda/FWHM${}, the central wavelength, and other filter parameters are given in Table~\ref{table:filters}.

\begin{table}[h!tbp]
\caption{Filter parameters}
\begin{center}
\begin{tabular}{ l | l | l}
\hline
Parameter & $L^\prime$ & $M^\prime$ \\
\hline
Central wavelength, $\mu$m & 3.77 & 4.68\\
FWHM, $\mu$m & 0.63 & 0.2\\
Cut-on (at 10\% of the peak), $\mu$m & 3.39 & 4.52\\
Cut-off (at 10\% of the peak), $\mu$m & 4.18 & 4.83\\
Average transmission, \% & 92 & 86\\
Out-of-band transmission, \% & <0.2 & <0.5\\
Substrate diameter, mm & \multicolumn{2}{c}{25}\\
Light diameter, mm & \multicolumn{2}{c}{20}\\
Thickness, mm & \multicolumn{2}{c}{3}\\
\hline
\end{tabular}
\label{table:filters}
\end{center}
\end{table}

For the possibility of manual calibration of the device, a hole is provided in the case, into which a cooled (for measuring dark current) or heated metal plate (for obtaining a map of pixel sensitivity distribution) can be inserted.

\begin{figure}[h]
\center{
\includegraphics[width=0.49\linewidth]{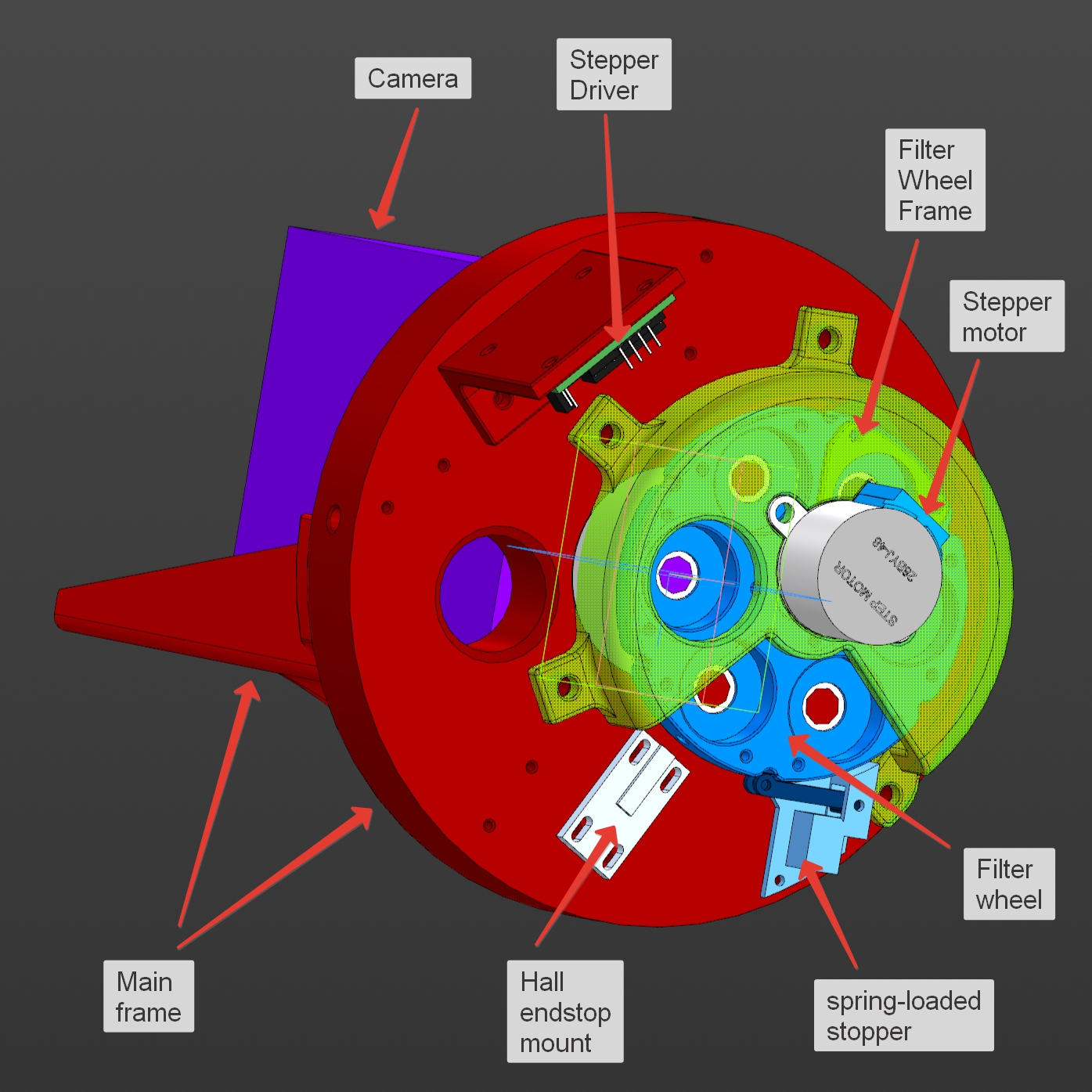}
\includegraphics[width=0.49\linewidth]{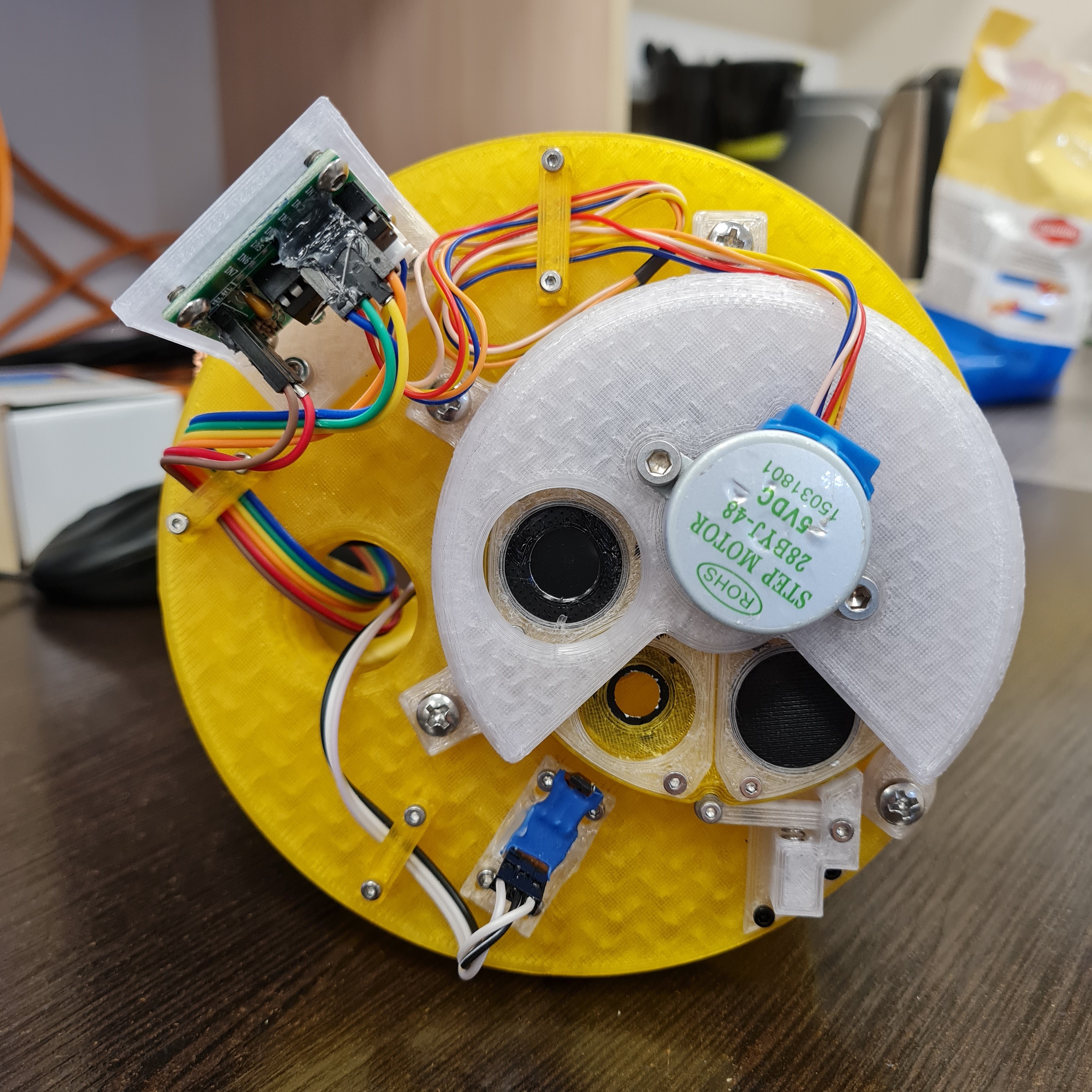}}
\caption{3D model of the photometer with filter wheel (left) and a photo of the finished device from the side of the filter wheel (right).}
\label{fig:lmp_photo}
\end{figure}

\begin{figure}[h]
\center{
\includegraphics[width=0.49\linewidth]{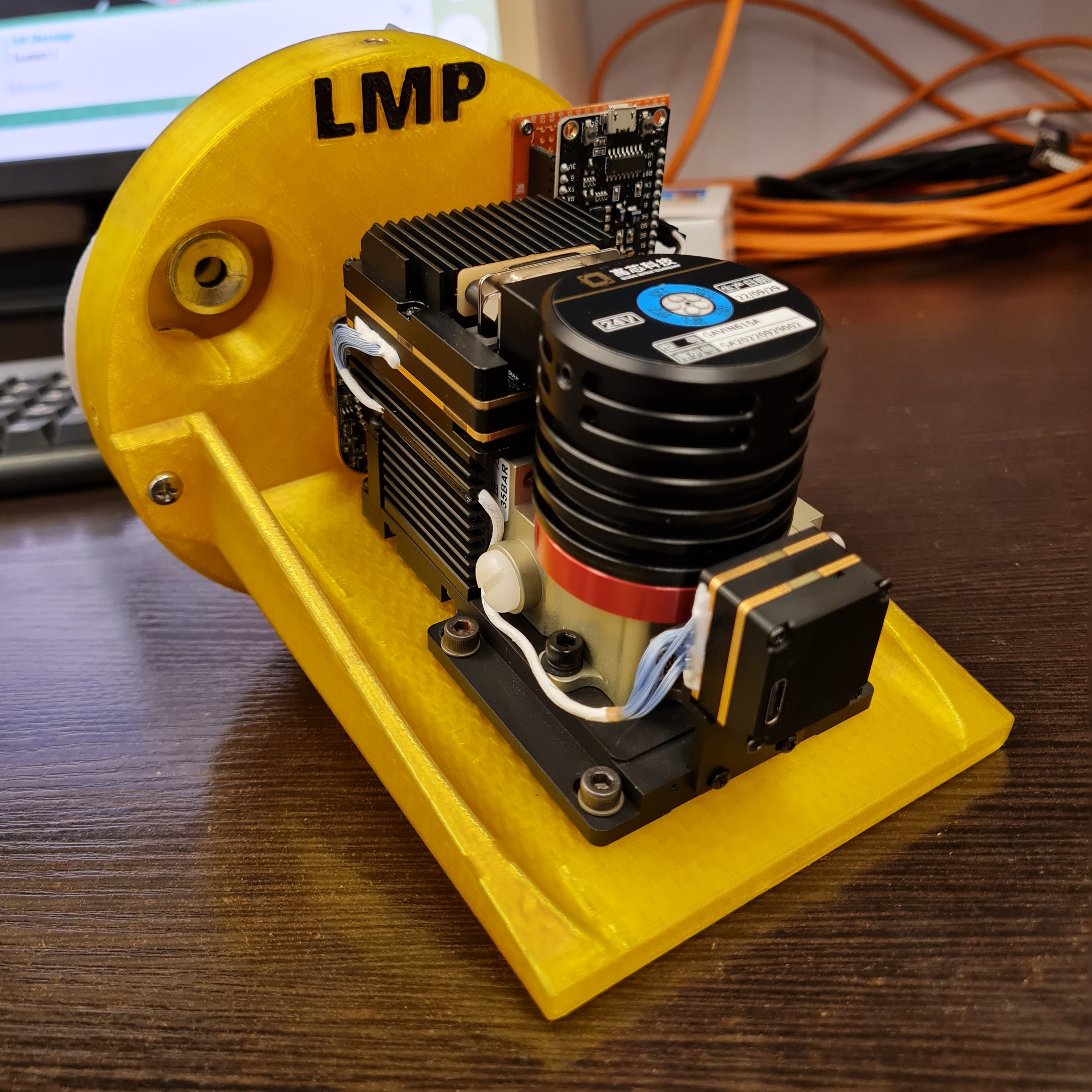}
\includegraphics[width=0.49\linewidth]{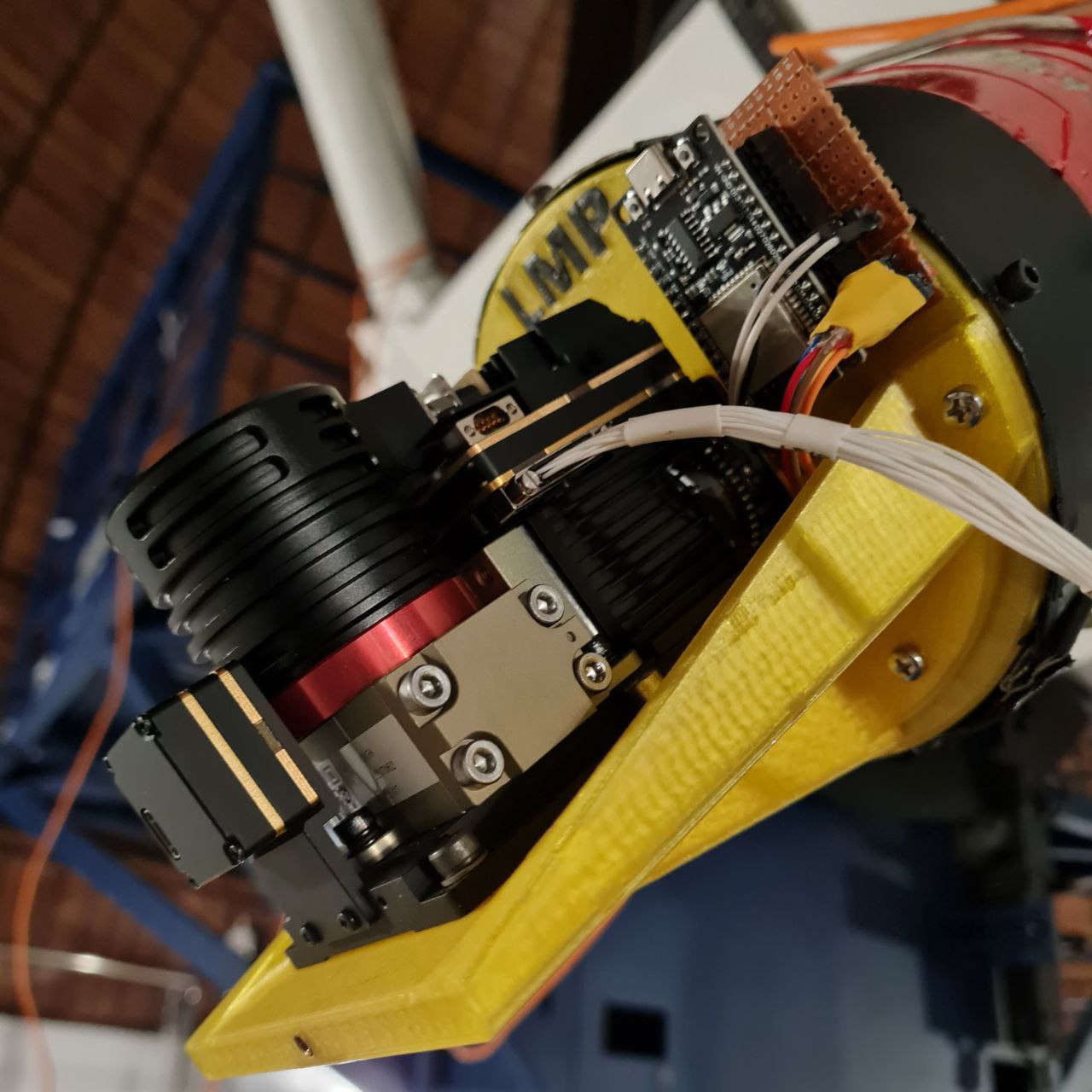}}
\caption{Photo of the finished device (left) and the photometer mounted on the telescope (right).}
\label{fig:lmp_photo2}
\end{figure}

All main parts of the device body and filter wheel are 3D-printed from PETG-plastic. Experience has shown that its stability is sufficient to ensure the required positioning accuracy of all photometer components. Fig.~\ref{fig:lmp_photo} and Fig.~\ref{fig:lmp_photo2} show the 3D model of the setup and its photos. As can be seen, the Gavin-615A photosensitive module is mounted on a monolithic part (shaded red in the 3D model) with a flange thickness of 20~mm and a mounting table thickness of 8~mm. The part has a 30\% fill rate when printed in 0.2~mm thick layers. The total weight of the photometer is 1350~g, of which 900~g is the photosensitive module.

Images are registered using a single-board computer Banana Pi BPI-M5, based on a 4-core Cortex-A55 processor under the Ubuntu 20.04. The computer is installed on the telescope near the camera, since the use of USB\,3.0 port for data transfer from the camera does not allow to place the computer far enough away. This choice of computer is due to the fact that from the manufacturer we managed to get the basic software (SDK) only for aarch64 architecture. On the basis of the SDK, which lacked some important functions, we developed our own software in C++ and Python. It can be used to acquire frames and control camera parameters through the built-in serial port. It is possible to control exposure in the range from 346 microseconds to 50 milliseconds, frame reading frequency, reading mode (raw frame, frame calibrated with internal algorithms of the module, color frame with superimposed thermal palette) and shutter trigger mode (automatic exposure start or using an external signal), as well as to get information about the state and temperature of the detector. On the basis of our improved SDK we wrote programs both for automatic acquisition of observing series and a graphical user interface for the operator of the 2.5-m telescope of CMO. The computer also controls the filter wheel and the instrument power supply. The observation control system with the LMP photometer is integrated into the \textit{Oplan} system for planning and controlling observations of the 2.5-m telescope.

\section{Observations}
\mbox{}\vspace{-\baselineskip}

Astronomical observations with the LMP photometer were carried out on the 2.5-m telescope of CMO. The telescope has one Cassegrain focus, two main Nesmith focuses (overlooking the balconies and equipped with derotators) and two additional Nesmith focuses, which are simply holes in the middle of the telescope tube. The LMP camera is mounted in one of the additional focuses. During the observations, low-frequency modulation of the light flux with an angular amplitude of $10''-15''$ for point objects and $1'$ for extended objects was performed by moving the telescope. Depending on the brightness of the object and the planned total duration of accumulation, the number of individual frames and the exposure time in each frame after moving the telescope tube were selected. The time taken by the telescope to move the telescope is practically independent of the displacement angle (within the angles used for modulation) and is 3 s.

To measure the brightness of the sky background, we took a photometer from a telescope. The photometer was mounted on a massive base so that its optical axis was positioned horizontally. A ZnSe lens with a focal length of 38 mm was installed in front of the entrance window with the filter. The sky radiation was guided into the lens using a flat mirror fixed on the axis of a stepper motor. By rotating the mirror, the field of view of the photometer was changed on the celestial sphere in such a way as to perform scanning from the zenith to the horizon in the selected vertical.

Most of the measurements used to obtain the parameters of the camera were made with the black body (BB) model made at the SAI many years ago. This device is a heater in the form of a spiral located on the perimeter of a metal cylinder. The cylinder is placed inside an insulating container. Asbestos cloth is used for thermal protection inside the container. Radiation from the heated cavity of the cylinder passes through a small hole in its base. In front of the hole, a rotating disk with $34$ holes of different diameters is mounted on heat-insulating legs, the use of which allows to vary the area of the radiating surface (the heat insulation eliminates the heating of the disk and reduces its own radiation of the diaphragms). The BB temperature is measured using a laboratory thermometer with a scale from $200$ to $300^{\circ}$C, which is placed in the cavity through a hole in the opposite base of the cylinder. Due to the large mass of the inner cylinder and high thermal inertia, the time to establish the operating temperature of the BB is about $5$ hours, which makes it possible to obtain high temperature stability in short time intervals required for measurements.

\mbox{}\vspace{-\baselineskip}

%%%%%%%%%%%%%%%%%%%%%%%%%%%%%%%%%%

\section{Camera characterization }
\mbox{}\vspace{-\baselineskip}

The basis of the LMP photometer is a light-sensitive module Gavin-615A manufactured by Global Sensor Technology (China) on the basis of HgCdTe 640x512/15$\mu$m MWIR detector and RS085 cryomodule. The manufacturer reports only some parameters of the detector (see Table~\ref{table:parameters}). In addition to these basic parameters, the GAIN, read noise, dark current and BIAS are required to process the observations.

\begin{table}[h!tbp]
\caption{Parameters of the Gavin-615A module installed in the LMP photometer (according to the manufacturer)}
\begin{center}
\begin{tabular}{ l | l }
\hline
Parameter & Value \\
\hline
Detector & HgCdTe \\
Format & 640x512 \\
Pixel size & 15x15 $\mu$m \\
Sensitivity spectral region (at 50\% level)  & 3.0 - 4.9 $\mu$m \\
Nonlinearity & < 8\% \\
Pixel capacity & $7.8\cdot10^6$\,e- \\
NETD & $\le 20$~mK (at 23$^\circ$C) \\
Discretization & 14 bit \\
Operating temperature of the detector & 85~K  \\
Cool down time & $\le 8$ minute  \\
\hline
\end{tabular}
\label{table:parameters}
\end{center}
\end{table}

\subsection{Nonlinearity}
\mbox{}\vspace{-\baselineskip}

Detector nonlinearity means the nonlinear dependence of the detector response, represented in relative digital ADU units and recorded in the final image file, on the magnitude of the incident radiation flux. For a constant radiation flux, this is equivalent to an accumulation time dependence.

\begin{figure}[h]
\center{
\includegraphics[width=0.9\linewidth]{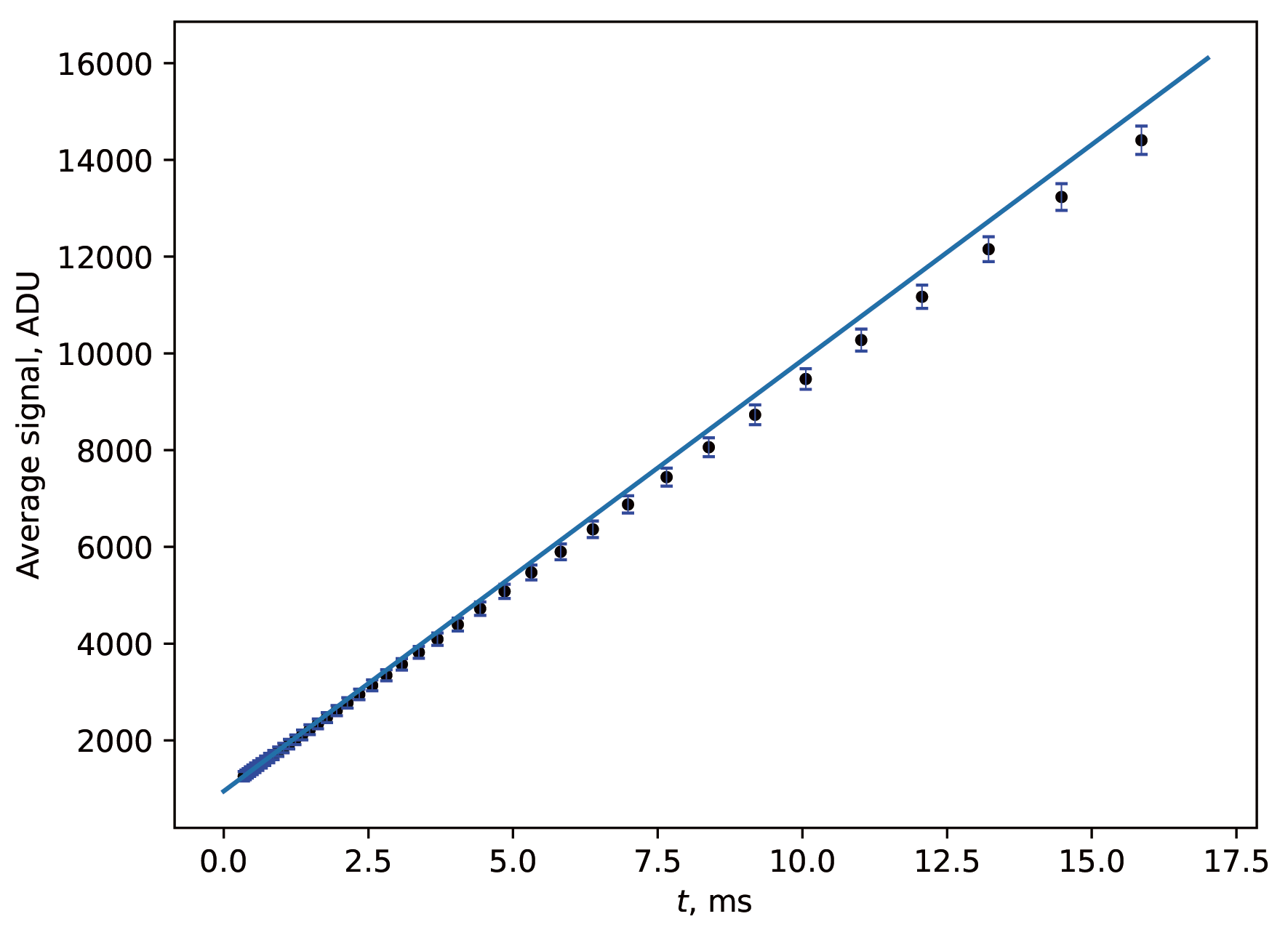}}
\caption{Dependence of the signal on the accumulation time $t$ (points) and the line $s(t)$ drawn along the four initial points}
\label{nelin0}
\end{figure}

We took a series of frames of the BB with different accumulation times: from the minimum $\approx$0.35~ms to $\approx$16~ms, when the filling of the cell capacitance occurred. Fig.~\ref{nelin0} shows a plot of the dependence on exposure $t$ of the average signal $F_{meas}$ in the center region of a 100x100 pixel frame. Assuming, as was done in \cite{Hilbert2008}, that in the first measurements the nonlinearity is small and comparable to the contribution of noise, we drew a straight line $s(t)$ along the four starting points to illustrate the expected signal values assuming linearity of the detector response. The monotonic deviation of the recorded signal from this straight line is clearly visible in the figure.

\begin{figure}[h]
\center{
\includegraphics[width=0.9\linewidth]{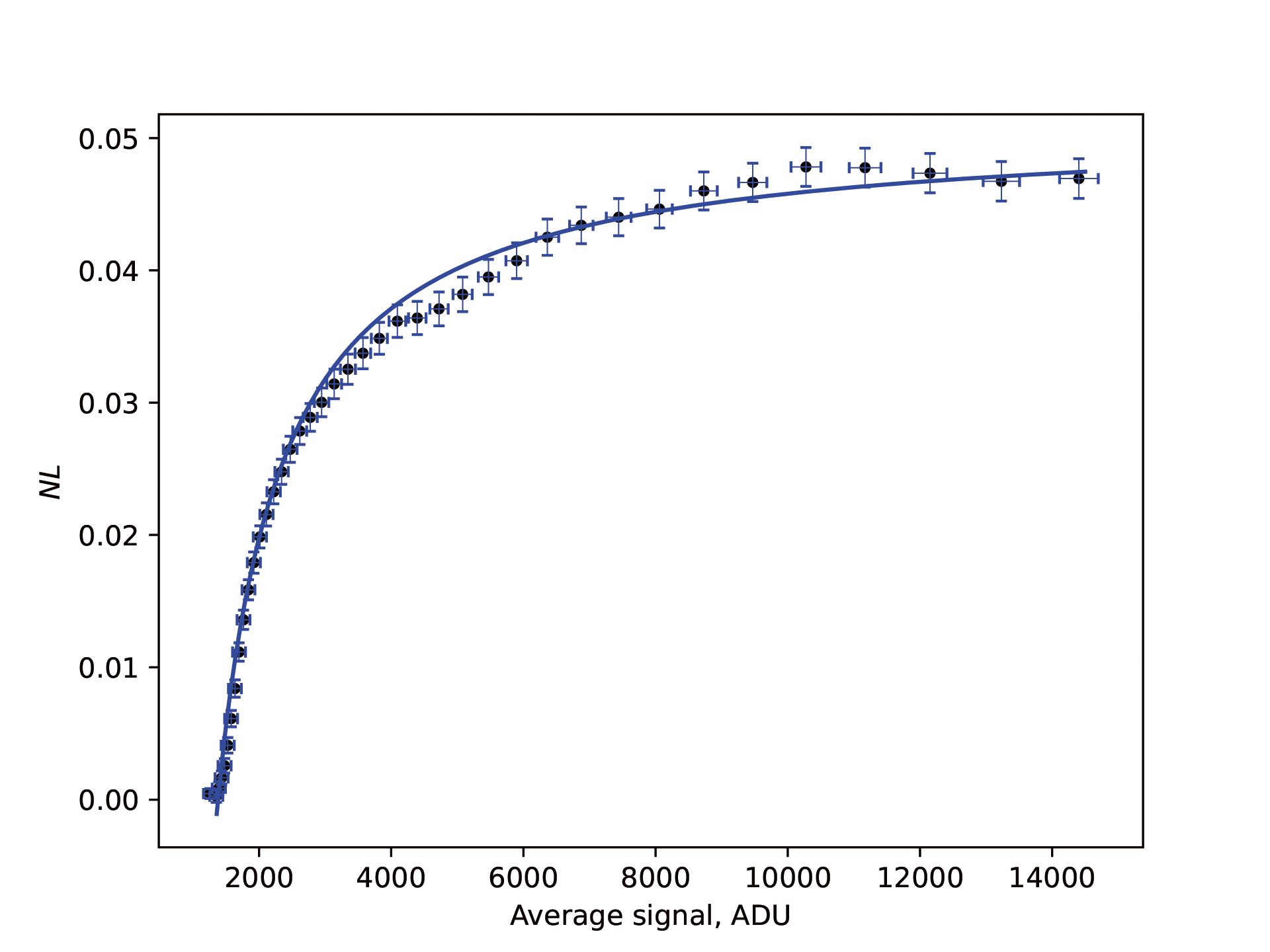}}
\caption{Dependence of the nonlinearity magnitude $NL=s(t)/F_{meas}-1$ on the mean signal (points) and its approximation by a hyperbola (solid line)}
\label{nelin}
\end{figure}

Then, according to the algorithm from \cite{Hilbert2008}, the values $s(t)/F_{meas}-1$ were calculated. Their dependence on the measured signal has a complex form (see Fig.~\ref{nelin}), which can be approximated by a polynomial of degree 6-7 $f(F_{meas})$ or by the function $f(F_{meas})=0.051-50/(F_{meas}-400)$. Knowing this function allows us to make a correction for nonlinearity: $F_{corr}=F_{meas}[1+f(F_{meas})]$. The degree of the polynomial turns out to be larger than in the \cite{Hilbert2008}, which suggests using the 3rd degree, but smaller than the 7-9 degree in the \cite{Chilingarian2014} pipeline. The dependence itself is also different~--- in \cite{Hilbert2008} it is a parabolic dependence, while in ours it is hyperbolic.

\subsection{Conversion coefficient}
\mbox{}\vspace{-\baselineskip}

The ratio between the digital data in ADU units and the number of photoelectrons captured by the detector is characterized by the conversion coefficient (GAIN), which is measured in [e$^-$/ADU].

\begin{figure}[h]
\center{
\includegraphics[width=0.85\linewidth]{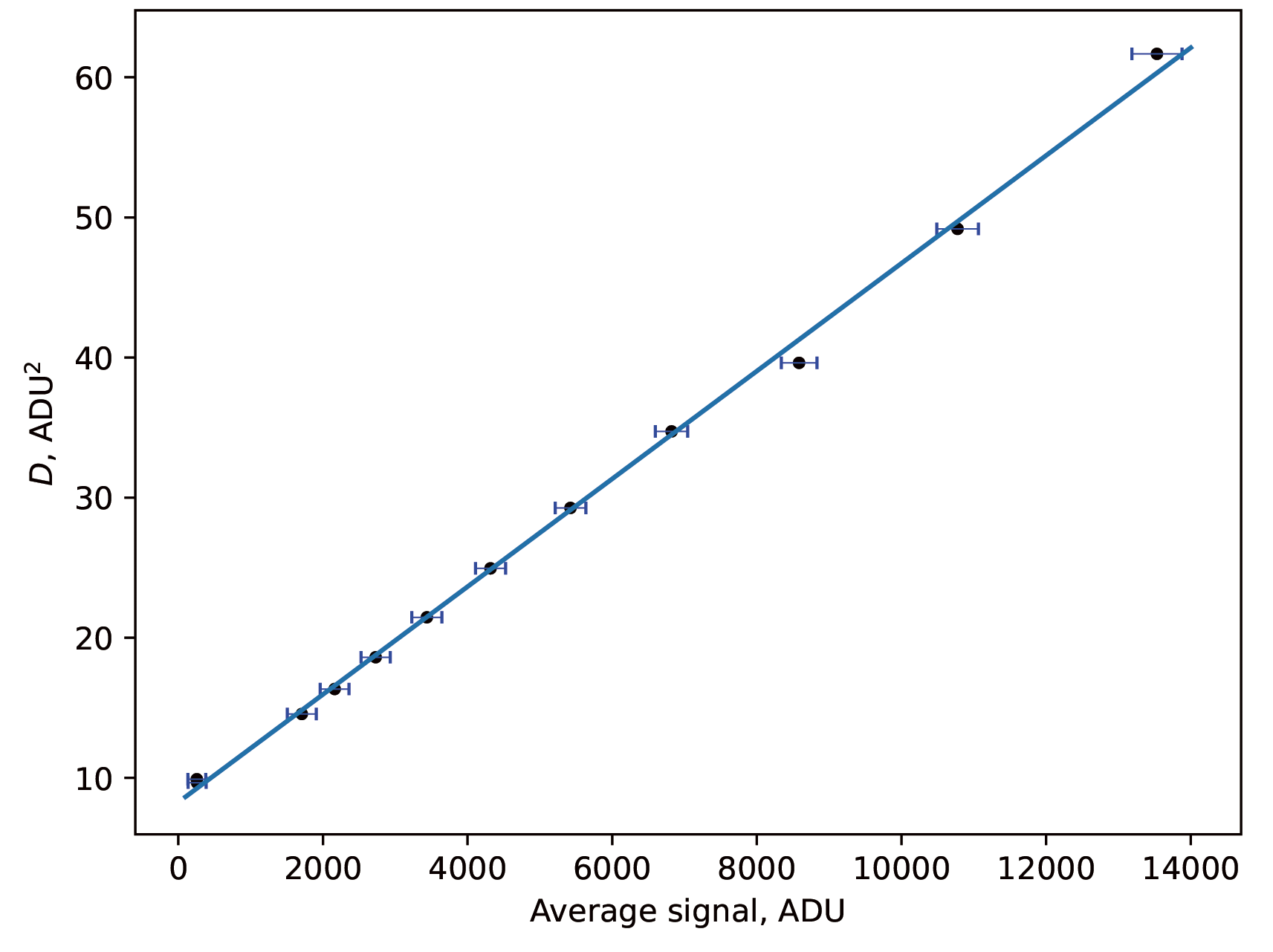}
}
\caption{Dependence of variance of difference frames on the average signal}
\label{gain}
\end{figure}

To measure the GAIN value, we obtained two series of 11 frames of the BB with pairwise equal accumulation times. The data were corrected for nonlinearity and the average signal level in ADU as well as the variance of the pairwise frame differences were calculated from them. The dependence of the signal variance on the average level was plotted from these series (Fig.~\ref{gain}). The GAIN value is defined as twice the inverse of the slope of the straight line approximating this dependence. For LMP camera, GAIN~= $520 \pm 9~e^{-}/$~ADU.

\subsection{Readout noise and bias frame}
\mbox{}\vspace{-\baselineskip}

Another important parameter of any detector that determines its ability to observe faint objects~--- readout noise (RN). It can be measured from a series of bias frames, which are usually recorded at zero exposure with the shutter closed. However, the light-sensitive module of an LMP photometer has no shutter (unlike CCD cameras), it cannot be set to zero exposure, and even in the dark it stores signal from ambient thermal radiation. Therefore, to record bias frames, we used the minimum available exposure and installed a metal plate cooled in liquid nitrogen in front of the entrance window of the module.

In this work, 4 series of 1000 frames with an accumulation time of 346 microseconds were obtained to estimate the readout noise and master-bias frame formation. From these series, the average frame (bias frame) and the distribution frame of the standard deviation of the recorded signal (i.e., the distribution frame of the read noise value) were calculated by pixel-by-pixel signal averaging. The frame-averaged RN~$=2.3\pm0.6$~ADU or, in terms of electrons, $1200\pm310$~e$^-$.

To determine bias, the same series with minimum accumulation time are used as for measuring read noise. Now on these series was calculated pixel-by-pixel average value, which is the bias value we need, because it is already said above that on these frames only it remains. The obtained BIAS~$= 960.5 \pm 2.2~ADU$.

As will be shown below, the detector has a high dark current. However, its noise at minimum exposure does not exceed $\sim$60~e$^-$ (0.1~ADU), and we neglected its contribution when calculating RN and BIAS.

\subsection{Dark current}
\mbox{}\vspace{-\baselineskip}

Another detector parameter can be a factor limiting sensitivity~--- dark current. It can be a major source of noise when large. Measurement of dark current in infrared detectors requires the use of a cold blanking plate placed in front of the detector's entrance window and placing the detector in a cooled enclosure to avoid reflection of radiation from the warm parts of the instrument from the blanking plate to the detector.

In our case, we used a blackened metal plate cooled in liquid nitrogen and located 5~mm from the entrance window of the photosensitive module. In images taken under these conditions, the signal consists of bias with readout noise, the dark signal accumulated over the exposure time, the intrinsic thermal emission from the entrance window of the detector, and the radiation reflected from the cold blanking plate, which always remains despite the protection methods taken. The entrance window of the module has an AR coating, so we will neglect its contribution (or rather, we will consider this contribution as a part of the dark current of the detector, depending on the external temperature). The radiation reflected from the cold plate due to the presence of a cold aperture inside the light-sensitive module falls only on the outer edges of the frame. That is, by examining the central part of the frame, it is possible to obtain an estimate of the dark current.

\begin{figure}[h]
\center{
\includegraphics[width=0.6\linewidth]{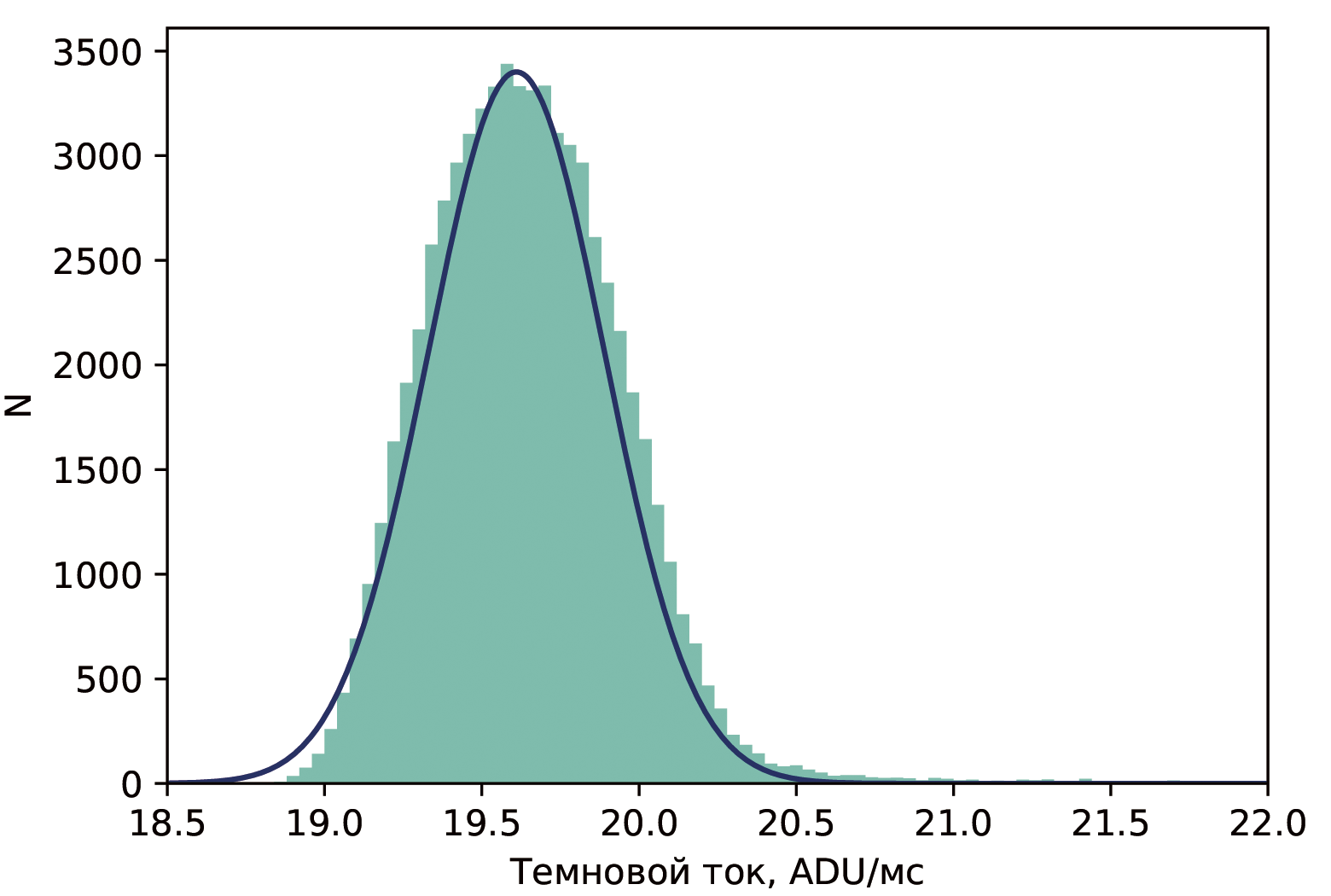}}
\caption{Histogram of dark signal magnitude in the central region of the detector and normal distribution curve with mean dark current 19.61 and $\sigma=0.28$}
\label{fig:ir_dark_hist}
\end{figure}

We performed 8 series of measurements of the dark current at an entrance window temperature of $\approx6^\circ$C. In each of them, the dark current was calculated for a frame area of 250x250 pixels using a series of frames obtained with exposures from minimum to 65~ms. For each pixel in this region, a linear approximation of the time dependence of the accumulated signal was performed. The histogram of the dark current for one of the series is shown in Fig.~\ref{fig:ir_dark_hist}. It can be seen that its shape is close to a normal distribution. This shows in favor of the correct cutting out of the area of the frame not illuminated by reflected radiation. The average value of the dark current measured in this series turns out to be 19.6~ADU/ms. The average over all series of the LMP-photometer dark current was found to be $17.8\pm2.2$~ADU/ms or $\approx(9.3\pm1.1)\cdot10^6$~e$^-$/s.

In addition to "normal"{} pixels whose dark signal accumulation rate corresponds to a normal distribution, located both in the region we studied and at the edges of the frame, there are $\approx$200 pixels (<0.1\% of the number of working pixels) whose dark current significantly ($>4\sigma$) exceeds the average value. These are so-called "hot"{} pixels characterized by an increased rate of "dark"{} electrons generation.

\subsection{First light and observations with LMP photometer}
\mbox{}\vspace{-\baselineskip}

\begin{figure}[h]
\center{
\includegraphics[width=0.3\linewidth]{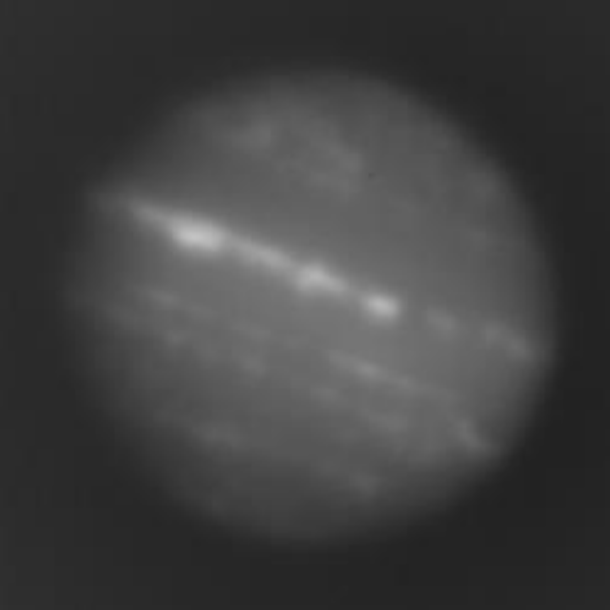}
\includegraphics[width=0.3\linewidth]{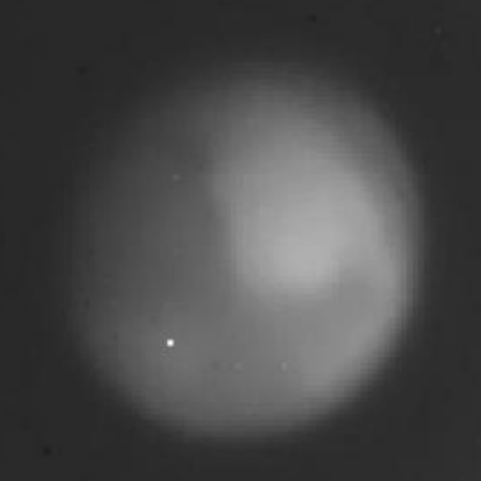}}
\caption{Photographs of Jupiter and Mars obtained on February 02, 2023 during photometer test. Angular sizes of planets at the moment of observations 36$''$ and 10.7$''$, respectively}
\label{fig:planets}
\end{figure}

\begin{figure}[h]
\center{
\includegraphics[width=1\linewidth]{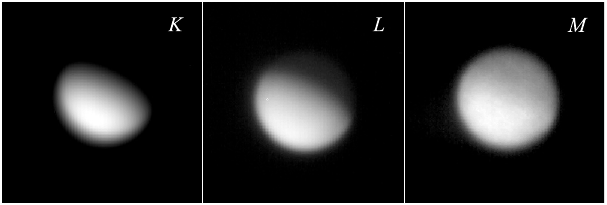}}
\caption{Images of Venus taken on April 13, 2023. Phase of Venus 0.75, angular diameter 14$''$}
\label{fig:venus}
\end{figure}

The first light with the prototype photometer was obtained on January 31, 2023 on the 2.5 meter telescope KGO GAISH MSU. The well-known carbon star CW~Leo was chosen as the object for the first observations as one of the brightest objects in the infrared sky. During these observations, the position of the focal plane was refined (the out-of-flange position of the Nasmyth N3 focus at the position of the secondary mirror corresponding to the sharp images in the other Nasmyth N1 focus, where the ASTRONIRCAM near-infrared camera was installed, was 102~cm), estimates of the background level, noise, typical exposure time, and FWHM of the stellar images were made, and images of Jupiter and Mars were obtained (Fig.~ \ref{fig:planets}). In April 2023, images of Venus were obtained in $K, L$, and $M$ filters, shown in Fig.~\ref{fig:venus}. It is interesting to note, despite the fact that the phase of the planet was 0.75 during the observations, the entire disk of Venus is visible in the $M$ band.

\begin{figure}[h]
\center{
\includegraphics[width=0.3\linewidth]{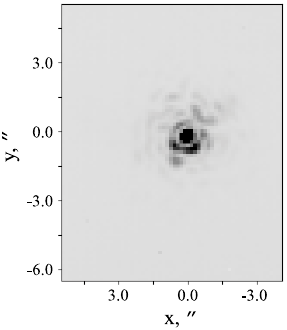}}
\caption{Brightness distribution in the image of the carbon star RW~LMi obtained on March 02, 2023 in the $M$ band}
\label{fig:star_profile}
\end{figure}

The prototype photometer has an unvignetted field of view of $30''$, with vignetting at 50\%~--- $50''$. The image scale is $\approx$$0.15''$/pixel. The typical sizes of the stellar images (FWHM ) in the averaged images are $\sim 0.6-0.7''$ (4-5 pixels), which is close enough to the diffraction limit for a wavelength of 4.8~$\mu$m ($0.48''$). Fig.~\ref{fig:star_profile} shows an example of an image of a star obtained under a quiet atmosphere with an exposure of 10~ms (single frame). The first diffraction ring and fragments of the following rings are clearly visible.

At the end of February 2023, after the installation of the filter wheel, observations of stars at different stages of evolution were started with the LMP photometer. For example, carbon stars V~Cyg, $\chi$~Cyg, S~Cep, T~Dra, young star ZZ~Tau IRS (\cite{Burlak2024}), as well as standard stars of different spectral types from the list of \cite{Shenavrin2011}. From the observations of the standards, the sensitivity of the instrument can be determined. At signal-to-noise ratio SNR=3 and good image quality, it is possible to observe stars up to 9.3$^m$ and 8.1$^m$ in the $L$ and $M$ bands, respectively, during the accumulation time of 20 s.

\begin{figure}[h]
\center{
\includegraphics[width=0.5\linewidth]{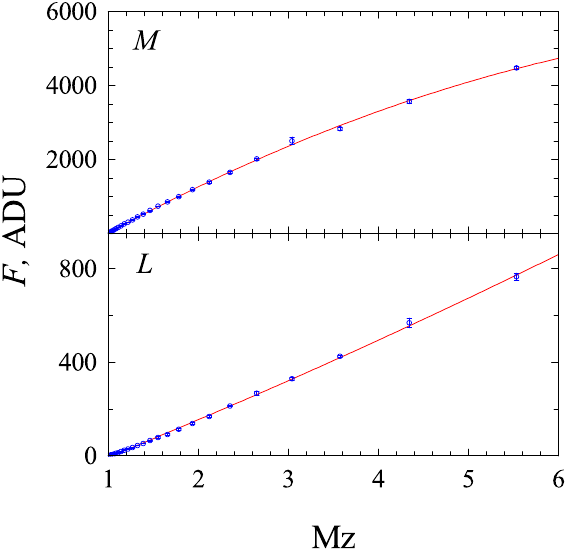}}
\caption{Dependence of sky brightness relative to zenith on air mass for filters $L$ and $M$}
\label{fig:airm}
\end{figure}

Since August 2023, the LMP photometer has also been used in the study of sky brightness at wavelengths of 3-5~µm. For this purpose, a separate setup has been built to provide sky scans in a selected vertical. The scanning step can be chosen to be any step, but $\sim5^\circ$ is usually used. After each step, measurements of brightness at zenith are made, which allows to compensate for fluctuations in the intrinsic radiation of the filters, lens and instrument walls on the assumption of constant sky brightness for short time intervals ($\sim$3 seconds).

An example of the sky brightness record obtained on October 20, 2023 at KGO is shown in Fig.~\ref{fig:airm}. The points on the graph represent the differences in signal from the background at airmass Mz and Mz=1 (at zenith) averaged over 10 separate scans. It can be seen that at small air masses the sky brightness depends almost linearly on Mz in both filters, following the expected dependence for spherically symmetric air layers. At Mz>2 in the $M$ band there is a marked difference from the straight line~--- the signal grows slower than expected.  We attribute this to the location of the setup on top of a hill, because of which, at about $>50$~m, the surface begins to deviate from the plane and the beam of view begins to cross higher ( colder and less humid) layers of the atmosphere than if we were observing over a flat surface. Such dependence is not observed in the $L$ band, which is due to its lower sensitivity to air temperature and humidity (this band has a shorter wavelength and the atmosphere is more transparent, see Fig.~\ref{table:filters}).

\section*{Conclutions}
\mbox{}\vspace{-\baselineskip}

The need to expand the spectral range, in which observations can be carried out on the 2.5-m telescope of the KGO, towards long wavelengths is beyond doubt~--- this is required by modern research trends and it is allowed by the astroclimate of the observatory and the appearance of an appropriate detector. At the same time, the possibilities of studying different classes of objects are significantly expanded: young stars, AGB and post-AGB stars, comets, bright AGN with excess IR radiation, and, finally, objects with circumstellar dust shells.

The photometer design we have implemented at this stage is not optimal in terms of minimizing the instrumental background and matching the scale to the average image quality in the CMO (instead of the recommended 3 pixels per FWHM of a point source, the $\approx5-6$ pixels is implemented). However, it allows, with a minimal project budget, to obtain a working tool that can access hundreds of thousands of objects on the sky.

The basis of the instrument is the light-sensitive module Gavin-615A, which was studied by us in laboratory conditions and in real conditions of the dome space. To ensure its operation and the possibility of interaction with the software and equipment of the 2.5-m telescope, we improved the software provided by the supplier and developed new ones that implement various modes of photometer operation.

The module detector has a large pixel capacity (see Tab.~\ref{table:parameters}), which, combined with the 14-bit digitization of the signal, results in a large conversion factor GAIN$ \approx 520$~e$^-$/ADU.

The nonlinearity of the detector turned out to be low, and according to our estimates at full pixel filling is $\le5$~\% (less than the manufacturer's stated value). The curve of nonlinearity dependence on the signal has a complex form~--- nonlinearity increases rapidly up to 4\% as the cell is filled, but after reaching 30\% of the capacity it almost stops changing. Correction functions with the coefficients we found can be used to account for it over the entire signal range.

Unfortunately, the detector has a large dark current (consisting of the detector's own dark current and radiation from the entrance window of the photosensitive module), which we estimate to be $\approx(9.3\pm1.1)\cdot10^6$~e$^-$/s at an entrance window temperature of $6^\circ$C. That is, the pixel is completely filled with dark signal in a time <1~s. Such exposure for the mid-infrared range is not a significant limitation, since due to the presence of a large instrumental and atmospheric background and a large flux of IR photons from objects, the typical accumulation time in IR instruments is usually tens of milliseconds.

When the signal is registered, the noise will be composed of the noise of the measured signal and detector noise, among which in our case the dark current noise and readout noise, which is $1200\pm310$~e$^-$, are dominant. These noises equalize at exposures of about 150~ms. At operating exposures of 10-20~ms, the dark current noise becomes 3-4 times less than the read noise.

Observations with the LMP photometer were started on the 2.5-m telescope, showing the possibility of using it to study objects up to 8-9 magnitude in the $L, M$ bands. Thus, we obtained estimates of the brightness of several bright carbon stars, standard stars of different spectral types from the list of \cite{Shenavrin2011} and a faint young star ZZ~Tau IRS (\cite{Burlak2024}), which had a brightness of $L=7.96$ and $M=6.78$ at the time of the observations, and for which the SNR$\sim10$ ratio was obtained for an accumulation time of 20~s (sum of 1000 frames of 20~ms). The estimation of the LMP photometer limiting agnitude shows that at SNR=3 and good image quality, it is possible to observe stars up to $\sim$9$^m$ and $\sim$8$^m$ in the $L$ and $M$ bands, respectively, during the accumulation time of 20 seconds.

At present, one of the factors limiting the photometer's ability to observe faint objects is the absence of a derotator at the focus of the Nasmyth N3 2.5-telescope. This leads to rotation of the field of view, which becomes noticeable already at measurement times longer than a few minutes. As a result, stacking individual frames to increase the signal-to-noise ratio becomes possible only when observing relatively bright objects brighter than approximately $7^m-8^m$, i.e., those objects whose images can be registered during the observation time (including accumulation time and overhead for signal reading and modulation) not exceeding several minutes. When observing extended objects, the brightness requirements increase even more. The installation of a mechanical derotator or the development of a software field rotation correction becomes the first priority to improve the efficiency of the photometer.
 
\bigskip

The authors are grateful for the support of the Development Program of M.~V.~Lomonosov Moscow State University (Scientific and Educational School <<Fundamental and Applied Space Research>>). The work of S.~Zheltoukhov (3D modeling, photometer assembly and adjustment, software development, observations) was supported by the Russian Science Foundation (RNF grant 21-12-00210). The authors are grateful to Vladimir Senyavin (Department of Physical Chemistry, Moscow State University) for obtaining the transmission curves of the filters.

% ======================================

\end {document}